\documentclass[conference]{IEEEtran}
\IEEEoverridecommandlockouts
\usepackage{cite}
\usepackage{amsmath,amssymb,amsfonts}
\usepackage{algorithmic}
\usepackage{graphicx}
\usepackage{textcomp}
\usepackage{xcolor}
\usepackage{graphicx}
\usepackage{subfigure}
\usepackage{verbatim}
\usepackage{balance}
\makeatletter
\renewcommand{\maketag@@@}[1]{\hbox{\m@th\normalsize\normalfont#1}}%
\makeatother
\newtheorem{definition}{Definition}
\def\BibTeX{{\rm B\kern-.05em{\sc i\kern-.025em b}\kern-.08em
    T\kern-.1667em\lower.7ex\hbox{E}\kern-.125emX}}
\begin{document}

\title{Protecting Personalized Trajectory with Differential Privacy under Temporal Correlations\\}
\author{
    \IEEEauthorblockN{Mingge Cao\IEEEauthorrefmark{1}, Haopeng Zhu\IEEEauthorrefmark{1}, Minghui Min\IEEEauthorrefmark{1}\IEEEauthorrefmark{2}, Yulu Li\IEEEauthorrefmark{1}, Shiyin Li\IEEEauthorrefmark{1}, Hongliang Zhang\IEEEauthorrefmark{3}, and Zhu Han\IEEEauthorrefmark{4}}
    \IEEEauthorblockA{\IEEEauthorrefmark{1} School of Information and Control Engineering, China University of Mining and Technology, Xuzhou 221116, China.}
    \IEEEauthorblockA{\IEEEauthorrefmark{2} Key Laboratory of Aerospace Information Security and Trusted Computing, Ministry of Education and School of\\ CyberScience and Engineering, Wuhan University, Wuhan 430072, China.}
    \IEEEauthorblockA{\IEEEauthorrefmark{3} School of Electronics, Peking University, Beijing 100871, China.}
    \IEEEauthorblockA{\IEEEauthorrefmark{4} Department of Electrical and Computer Engineering, University of Houston, Houston, TX 77004, USA.}
    \thanks{ \emph{This research was funded by the National Natural Science Foundation of China (grant number 62101557 and 62371451), and Xuzhou Basic Research Plan Project-Young Scientific and Technological Talent Project (KC23022), and China Postdoctoral Science Foundation (2022M713378), and the Fundamental Research Funds for the Central Universities (2042022kf0021), and NSF CNS-2107216, CNS-2128368, CMMI-2222810, ECCS-2302469, US Department of Transportation, Toyota and Amazon}. \emph{Corresponding author: Minghui Min (minmh@cumt.edu.cn)}.}}

\maketitle

\begin{abstract}
Location-based services (LBSs) in vehicular ad hoc networks (VANETs) offer users numerous conveniences. However, the extensive use of LBSs raises concerns about the privacy of users' trajectories, as adversaries can exploit temporal correlations between different locations to extract personal information. Additionally, users have varying privacy requirements depending on the time and location. To address these issues, this paper proposes a personalized trajectory privacy protection mechanism (PTPPM). This mechanism first uses the temporal correlation between trajectory locations to determine the possible location set for each time instant. We identify a protection location set (PLS) for each location by employing the Hilbert curve-based minimum distance search algorithm. This approach incorporates the complementary features of geo-indistinguishability and distortion privacy. We put forth a novel Permute-and-Flip mechanism for location perturbation, which maps its initial application in data publishing privacy protection to a location perturbation mechanism. This mechanism generates fake locations with smaller perturbation distances while improving the balance between privacy and quality of service (QoS). Simulation results show that our mechanism outperforms the benchmark by providing enhanced privacy protection while meeting user's QoS requirements.
\end{abstract}


\section{Introduction}
Location-based services (LBSs) in vehicular ad hoc networks (VANETs), such as real-time traffic information reports and personalized navigation, significantly enhance our daily lives \cite{10177933}, \cite{wang2019protecting}. However, to enjoy these convenient services, VANET users must  provide their real-time location to the LBS server, raising concerns about privacy breaches \cite{min2021reinforcement}. Several location privacy protection mechanisms have been developed to address this issue. However, focusing solely on protecting location information is insufficient. Trajectory data, which consists of interconnected locations, holds valuable temporal information that potential attackers can exploit to deduce users' activities and uncover sensitive personal information~\cite{tang2021dlp}. Besides, different users may have different location privacy and quality of service (QoS) requirements \cite{min20213d}, and even the same user may have various sensitive information at different times and locations, and thus have different privacy protection demands. Therefore, ensuring the privacy of user trajectories and meeting their personalized demands is of utmost importance.

Most existing research focuses on privacy protection for individual locations. For instance, a privacy notion called geo-indistinguishability, based on differential privacy, is proposed in \cite{andres2013geo}. This notion aims to protect a user's location within a certain radius, guaranteeing “generalized differential privacy”. However, this approach overlooks arbitrary prior knowledge that adversaries may possess, leading to potential privacy leakage \cite{shokri2012protecting}, and the degree of privacy protection is not clearly defined. To address these shortcomings, authors in~\cite{yu2017dynamic} combine geo-indistinguishability and expected inference error, leveraging their complementary properties to propose a personalized location privacy protection mechanism. However, this approach still fails to consider the temporal correlation between different locations on a trajectory.

A solution called “$\delta$-location set” based differential privacy is proposed in \cite{xiao2015protecting}, which combines the location privacy protection mechanism in \cite{yu2017dynamic} with the temporal correlation between locations on a trajectory. However, this approach assigns the same privacy budget to all locations and does not cater to users' personalized demands. A few trajectory privacy protection schemes use $k$-anonymity \cite{xing2021location} to achieve privacy protection. These methods generalize and aggregate individual trajectory data, ensuring the trajectory remains protected while combining it with at least $k-1$ other trajectories to form an anonymous region. However, these approaches rely on a trusted third party and fail to provide strict privacy guarantees \cite{jin2022survey}, \cite{wu2023tcpp}. In summary, existing trajectory privacy protection schemes  lack consideration of crucial aspects  such as the temporal correlation between locations on a trajectory, meeting the user's personalized needs, and ensuring the protection of the user's actual location without relying on a trusted third party. Consequently, a novel trajectory privacy mechanism is needed to simultaneously satisfy these requirements.

In this paper, we develop a personalized trajectory privacy protection mechanism (PTPPM) that considers the temporal correlation between locations on a trajectory. The mechanism constructs a location transition probability matrix, deriving the potential location set for the user at each time point along the trajectory. To improve privacy, we leverage the complementary features of geo-indistinguishability \cite{andres2013geo} and distortion privacy \cite{shokri2011quantifying} by employing the Hilbert curve-based minimum distance search algorithm \cite{yu2017dynamic} to identify a protection location set (PLS) encompassing all potential locations along the trajectory. Geo-indistinguishability can limit the attacker's posterior knowledge, but cannot quantify the similarity between the attacker's inferred location and the actual location. Distortion privacy can ensure that the attacker's expected inference error is greater than a certain threshold. However, it cannot prevent the leakage of posterior information. The combination of these two notions can effectively strengthen the resistance against location inference attacks. The mechanism also enables personalized user privacy protection by adjusting the privacy settings through two privacy parameters.

In addition, we introduce an extension of the Permute-and-Flip mechanism \cite{mckenna2020permute}, originally designed for data privacy protection during data publishing, to serve as a location perturbation mechanism. This novel approach achieves a smaller perturbation distance, which has a better balance between location privacy and QoS. Simulation results demonstrate that PTPPM provides personalized trajectory privacy protection and offers superior privacy preservation compared to PIVE~\cite{yu2017dynamic} under the same QoS loss. The main contributions of our work include:
\begin{enumerate}
\item We propose a personalized trajectory privacy protection mechanism called PTPPM, which can defend the attacker that obtains the temporal correlation between various locations within a trajectory. This mechanism combines two privacy notions of geo-indistinguishability and distortion privacy to enhance the system's robustness against location inference attacks.

\item We put forth a novel location perturbation mechanism, Permute-and-Flip. It has a smaller perturbation distance to release perturbed locations, thereby achieving a better balance between location privacy and QoS.

\item We conduct comprehensive simulations to study the impact of different privacy budgets and expected inference errors on users' personalized requirements. Additionally, we demonstrate the performance advantage of PTPPM over PIVE under the same QoS loss.
\end{enumerate}

The remainder of this paper is organized as follows. Section \ref{System Model} presents the system model. We present the trajectory privacy protection statement in Section \ref{Trajectory Privacy Protection Statement}. A PTPPM framework is proposed in Section \ref{Presonalized Trajectory Privacy Protection Mechanism}. The evaluation results are provided in Section \ref{Simulation Results}. Finally, we conclude this work in Section \ref{Conclusion}.

\section{System Model}\label{System Model}
\begin{figure}[!t]

\begin{center}

\includegraphics[width=0.37\textwidth]{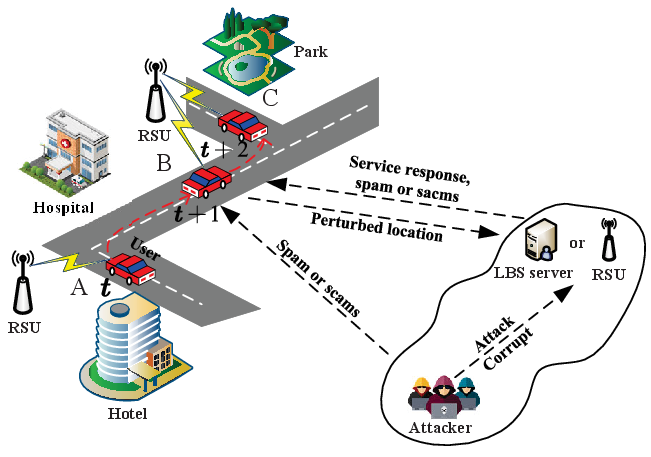}

\end{center}

\caption{Illustration of the trajectory privacy protection.}

\label{fig1}

\end{figure}

To obtain real-time LBSs, we consider that VANET users share their location information with a roadside unit (RSU) or an LBS server at different times and locations \cite{min20213d}, \cite{9201413}. Users interact with the RSU to access road information, plan their destinations, and determine driving routes. Fig. \ref{fig1} illustrates the user's driving trajectory, where A, B, and C are the user's locations at different times. To protect privacy, the user releases the perturbed locations. The untrusted LBS server, which an external attacker might also corrupt, can infer the user's sensitive information (e.g., the user's driving trajectory) at a particular time by analyzing the received temporally correlated location information and sending related spam or scams while providing feedback services.
\subsection{User Model}
We consider VANET users driving within a specific area of a city, which is divided into multiple grids. Each grid cell represents a distant location state of the user, and each cell is associated with a unique 2D coordinate. The state of all locations of the user in the area is $\mathcal{A}=\left\{ \boldsymbol{a}_1,\ \boldsymbol{a}_2,\ \cdot \cdot \cdot ,\ \boldsymbol{a}_n \right\}$, where $n$ is the total number of location states. $\boldsymbol{x}_t$ represents the user's true location at time $t$, and $\boldsymbol{l}_t$ represents the two-dimensional coordinates of the user's location state at time $t$. For example, a shown in Fig. \ref{User map coordinates and status coordinates}, $\mathcal{A}=\left\{ \boldsymbol{a}_1,\ \boldsymbol{a}_2,\ \cdot \cdot \cdot ,\ \boldsymbol{a}_{22} \right\}$, $\boldsymbol{x}_t=\boldsymbol{a}_6=\left[ 0,0,0,0,0,1,0 \cdot \cdot \cdot ,0 \right]$, $\boldsymbol{l}_t=\left[ 2,4 \right]$.

The user uses the location perturbation mechanism to remap the actual location $\boldsymbol{x}_t$ from the actual location set $O_1$ to the fake location $\boldsymbol{x}_{t}^{'}$ from the perturbed location set $O_2$. The location perturbation probability distribution $f$ is given by
\begin{equation}\label{2}
f\left( \boldsymbol{x}_{t}^{'}|\boldsymbol{x}_t \right) =\text{Pr}\left( O_2=\boldsymbol{x}_{t}^{'}|O_1=\boldsymbol{x}_t \right) ,\ \ \ \ \boldsymbol{x}_t,\boldsymbol{x}_{t}^{'}\in \mathcal{A}.
\end{equation}
We use $\boldsymbol{p}_t$ to represent the user's location state at time $t$, where $\boldsymbol{p}_t\left[ i \right] =\text{Pr}\left( \boldsymbol{x}_t=\boldsymbol{a}_i \right) =\text{Pr}\left( \boldsymbol{l}_t \right)$ represents the probability that the user's real location is in $\boldsymbol{a}_i$ at time $t$. Assuming that users are distributed with the same probability $\mathcal{A}=\left\{ \boldsymbol{a}_2,\ \boldsymbol{a}_3,\ \boldsymbol{a}_5,\ \boldsymbol{a}_7 \right\}$, then the location probability distribution of users is $\boldsymbol{p}_t=\left[ 0,0.25,0.25,0,0.25,0,0.25,0,\cdot \cdot \cdot ,0 \right]$. We use $\boldsymbol{p}_{t}^{-}$ and $\boldsymbol{p}_{t}^{+}$ to represent the prior and posterior probabilities of the user before and after observing the released perturbed location $\boldsymbol{x}_{t}^{'}$.
\begin{figure}[htbp]

\begin{center}

\includegraphics[width=0.35\textwidth]{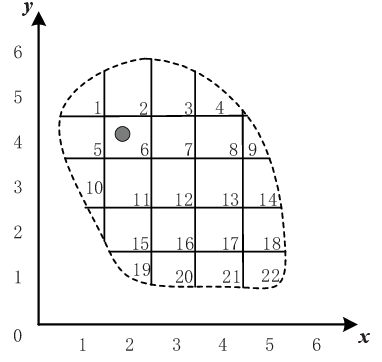}

\end{center}

\caption{User map coordinates and status coordinates.}

\label{fig2}

\end{figure}
\subsection{Attack Model}
We consider the attacker to be an untrusted LBS server or an external attacker who may attack or corrupt the LBS server. They can access users’ current location information for commercial profit or illegal purposes. We assume that the attacker knows the location perturbation probability distribution $f\left( \boldsymbol{x}_{t}^{'}|\boldsymbol{x}_t \right)$,
and can obtain the prior distribution $\boldsymbol{p}_{t}^{-}=\text{Pr}\left( \boldsymbol{x}_t \right)$ of the user’s current location through public tracking, check-in data set, or statistical information \cite{chatzikokolakis2015constructing}. Then, the attacker can calculate the posterior probability distribution $\boldsymbol{p}_{t}^{+}=\text{Pr}\left( \boldsymbol{x}_t|\boldsymbol{x}_{t}^{'} \right)$
after observing the user’s reported location $\boldsymbol{x}_{t}^{'}$, i.e.,
\begin{align}\label{Posterior probability}
\boldsymbol{p}_{t}^{+}=\text{Pr}\left( \boldsymbol{x}_t|\boldsymbol{x}_{t}^{'} \right) =\frac{\text{Pr}\left( \boldsymbol{x}_t \right) f\left( \boldsymbol{x}_{t}^{'}|\boldsymbol{x}_t \right)}{\sum_{\boldsymbol{x}_t\in \mathcal{A}}{\text{Pr}\left( \boldsymbol{x}_t \right) f\left( \boldsymbol{x}_{t}^{'}|\boldsymbol{x}_t \right)}}.
\end{align}

A Bayesian adversary aims to infer the actual location at time $t$ by minimizing the expected inference error against the posterior distribution. Therefore, the inferred location $\boldsymbol{\hat{x}}_t$ is
\begin{align}\label{inference error against}
\boldsymbol{\hat{x}}_t=\underset{\boldsymbol{\hat{x}}_t\in \mathcal{A}}{\text{arg}\min}\sum_{\boldsymbol{x}_t\in \mathcal{A}}{\text{Pr}\left( \boldsymbol{x}_t|\boldsymbol{x}_{t}^{'} \right)}d\left( \boldsymbol{\hat{x}}_t,\boldsymbol{x}_t \right).
\end{align}

We assume that the attacker can obtain the user's location transition probability matrix $\mathbf{M}$ based on the user's historical trajectory data and behavior habits \cite{xiao2015protecting}. Then, the attacker can infer the prior probability $\boldsymbol{p}_{t+1}^{-}$ of the user at time $t+1$, i.e.,
\begin{align}\label{t+1 prior probability}
\boldsymbol{p}_{t+1}^{-}=\boldsymbol{p}_{t}^{+}\mathbf{M}.
\end{align}

Since the posterior probability $\boldsymbol{p}_{t+1}^{+}$ at time $t+1$ can be obtained according to (\ref{Posterior probability}), the attacker can perform an optimal inference attack on the user's location at time $t+1$ according to (\ref{inference error against}) to obtain the corresponding inferred location $\boldsymbol{\hat{x}}_{t+1}$. Therefore, the attacker can obtain the inferred trajectory of the user in a certain period of time by performing the optimal inference attack on the user's location at each moment on the trajectory, thereby stealing the user's trajectory privacy.

\section{Trajectory Privacy Protection Statement}\label{Trajectory Privacy Protection Statement}
In this section, we first list the main trajectory privacy notions and the condition for determining PLS, and then we present this paper's problem statement.

\subsection{Location Transition Probability Matrix}
Matrix $\mathbf{N}$ is the location transfer matrix, representing the number of times a user goes from one place to another. Let $n_{ij}$ be an element in the $i$th row and $j$th column of matrix $\mathbf{N}$, and $n_{ij}$ represents the number of times the user goes from region $\boldsymbol{a}_{i}$ to region $\boldsymbol{a}_{j}$.

Through the location transition matrix $\mathbf{N}$, the location transition probability matrix $\mathbf{M}$ of the user can be analyzed. Let $m_{ij}$ be an element in the $i$th row and the $j$th column of matrix $\mathbf{M}$, $m_{ij}=\dfrac{n_{ij}}{\sum_j{n_{ij}}}$ represents the probability of the user moving from $\boldsymbol{a}_{i}$ to $\boldsymbol{a}_{j}$. The matrix $\mathbf{M}$ describes the temporal correlation of the user at different locations in a trajectory.
\subsection{$\delta$-Location Set}
To protect locations frequently visited by users, $\delta$-location set is proposed in \cite{xiao2015protecting}, which represents the set of locations where the user is most likely to appear at time $t$, and we denote it as $\varDelta \boldsymbol{\chi }_t$.

$\varDelta \boldsymbol{\chi }_t$ denotes a set containing the minimum number of locations at time $t$ with a prior probability sum not less than $1-\delta $ ($0<\delta <1$).
\begin{equation}\label{18}
\varDelta \boldsymbol{\chi }_t=\min\left\{ \boldsymbol{a}_i|\sum_{\boldsymbol{a}_i}{\boldsymbol{p}_{t}^{-}\left[ i \right]}\ge 1-\delta \right\}.
\end{equation}
Note that since the $\delta$-location set represents a set of possible locations with a high probability of the user appearing at time $t$, the real location $\boldsymbol{x}_t$ of the user may be eliminated with an extremely small probability. In this case, we substitute the closest location $\boldsymbol{\tilde{x}}_t$ for the actual location $\boldsymbol{x}_t$, given by
\begin{equation}\label{19}
\boldsymbol{\tilde{x}}_t=\underset{\boldsymbol{\tilde{x}}_t\in \varDelta \boldsymbol{\chi }_t}{\text{arg}\min}\ d\left( \boldsymbol{\tilde{x}}_t,\boldsymbol{x}_t \right).
\end{equation}
If $\boldsymbol{x}_t\in \varDelta \boldsymbol{\chi }_t$, then $\boldsymbol{x}_t$ is protected in $\varDelta \boldsymbol{\chi }_t$; if not, $\boldsymbol{\tilde{x}}_t$ is protected in $\varDelta \boldsymbol{\chi }_t$.

\subsection{Condition for Determining PLS}

A two-phase dynamic differential location privacy framework PIVE was proposed in \cite{yu2017dynamic}. It studies the complementary relationship between geo-indistinguishability and distortion privacy and obtains the upper bound of posterior probability and lower bound of inference error through formula derivation. By combining these two privacy notions, PIVE introduces a user-defined inference error bound $E_m$ to determine PLS.

First, to guarantee the expected inference error in terms of PLS, the conditional expected inference error is given by
\begin{equation}
ExpEr\left( \boldsymbol{x}_{t}^{'} \right) =\underset{\boldsymbol{\hat{x}}_t\in \mathcal{A}}{\min}\sum_{\boldsymbol{x}_t\in \mathcal{A}}{\text{Pr}\left( \boldsymbol{x}_t|\boldsymbol{x}_{t}^{'} \right)}d\left( \boldsymbol{\hat{x}}_t,\boldsymbol{x}_t \right).
\end{equation}
Given that the adversary narrows possible guesses to the PLS $\boldsymbol{\Phi }_t$ that contains the user’s true location, we define
\begin{equation}
E\left( \boldsymbol{\Phi }_t \right) =\underset{\boldsymbol{\hat{x}}_t\in \mathcal{A}}{\min}\sum_{\boldsymbol{x}_t\in \boldsymbol{\Phi }_t}{\frac{\text{Pr}\left( \boldsymbol{x}_t \right)}{\sum_{\boldsymbol{y}_t\in \boldsymbol{\Phi }_t}{\text{Pr}\left( \boldsymbol{y}_t \right)}}}d\left( \boldsymbol{\hat{x}}_t,\boldsymbol{x}_t \right).
\end{equation}

According to the lower bound on expected inference error,
\begin{equation}\label{22}
ExpEr\left( \boldsymbol{x}_{t}^{'} \right) \ge e^{-\epsilon}E\left( \boldsymbol{\Phi }_t \right),
\end{equation}
the authors in \cite{yu2017dynamic} (Theorem 1) obtain a sufficient condition,
\begin{equation}\label{Constraint condition}
E\left( \boldsymbol{\Phi }_t \right) \ge e^{\epsilon}E_m,
\end{equation}
to satisfy the user-defined threshold, $\forall \boldsymbol{x}_{t}^{'}$, $ExpEr\left( \boldsymbol{x}_{t}^{'} \right) \ge E_m$.
\subsection{Problem Statement}
Considering the temporal correlation between locations on the trajectory, it is insufficient to protect only the user's current location, as attackers can still deduce the actual location by analyzing behavior patterns, geographical constraints, and other available information. Assuming the attacker possesses knowledge of the user's location transition probability matrix $\mathbf{M}$, they can calculate the prior probability of the user's current location based on previously published location information. To enhance the protection of the user's current location, we focus on protecting frequently visited locations with high prior probabilities \cite{xiao2015protecting}.

Moreover, different types of LBS and varying contexts may impose different users' privacy requirements. Even for the same LBS, users may have various privacy needs for the same location at different times or in different locations. Therefore, we determine the possible location set $\boldsymbol{\Delta \chi }_t$ for each user at any given time based on the prior probability of locations along the trajectory. By combining the concepts of geo-indistinguishability and distortion privacy, we use a Hilbert-based minimum distance search algorithm to identify  the set $\boldsymbol{\Phi }_t$ of possible locations in $\boldsymbol{\Delta \chi }_t$ at any location along the trajectory. We personalize user privacy by adjusting the privacy budget $\epsilon$ and the expected inference error threshold $E_m$. To enhance performance, we adapt the Permute-and-Flip mechanism, which was originally designed for data publishing scenarios, to serve as the location perturbation mechanism.

\section{Presonalized Trajectory Privacy Protection Mechanism}\label{Presonalized Trajectory Privacy Protection Mechanism}
\begin{figure}[!t]

\begin{center}

\includegraphics[width=0.5\textwidth]{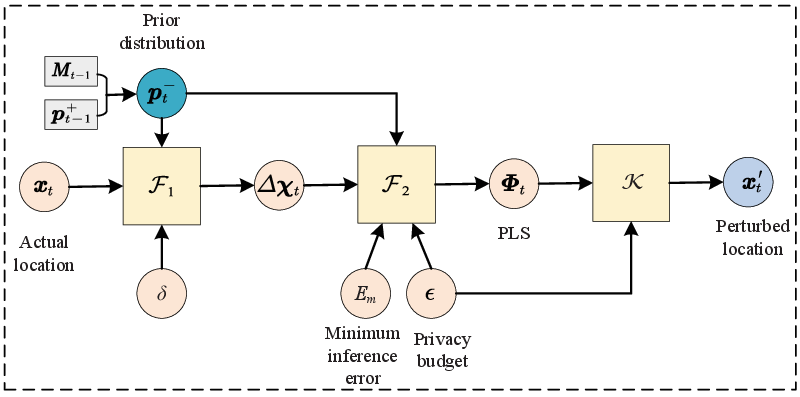}

\end{center}

\caption{The framework of PTPPM.}

\label{fig3}

\end{figure}

In this section, we propose a personalized trajectory privacy protection mechanism PTPPM, as shown in Fig. \ref{fig3}. Here, we consider the temporal correlation between different locations on the trajectory and combine geo-indistinguishability and distortion privacy to protect the user's personalized trajectory privacy. Specifically, we first use algorithm $\mathcal{F}_1$ to obtain the set of possible locations for the user at each time, leveraging the associated prior probability at each moment along the trajectory. Second, we employ algorithm $\mathcal{F}_2$ to dynamically select PLS for each possible location on the trajectory, incorporating both geo-indistinguishability and distortion privacy. Furthermore, our mechanism enables personalized trajectory privacy protection by adjusting different privacy settings (minimum inference error and privacy budget) for individual users. Finally, we put forth a novel Permute-and-Flip mechanism $\mathcal{K}$ to generate a perturbed location $\boldsymbol{x}_{t}^{'}$ for each location within the PLS. These perturbed locations are selected with a smaller perturbation distance, ensuring a better QoS experience while providing robust and effective privacy protection.

\subsection{Determine $\varDelta \boldsymbol{\chi }_t$ at Continuous  Times}
The transition probability matrix $\mathbf{M}$ is constructed according to the user's historical trajectory data and behavior habits \cite{xiao2015protecting}.
We eliminate all impossible locations ($\boldsymbol{p}_{t}^{-}$ is minimal or $\boldsymbol{p}_{t}^{-}=0$) based on certain criteria to obtain the set of possible locations at time $t$, i.e., $\varDelta \boldsymbol{\chi }_t$.
If the actual location at time $t$ is removed, we substitute it with $\boldsymbol{\tilde{x}}_t$.

%
%

We calculate the posterior probability $\boldsymbol{p}_{t}^{+}$ according to (\ref{Posterior probability}) and then combine the location transition probability matrix $\mathbf{M}$ according to (\ref{t+1 prior probability}) to obtain the prior probability $\boldsymbol{p}_{t+1}^{-}$ at time $t+1$. In terms of $\boldsymbol{p}_{t+1}^{-}$, we get $\varDelta \boldsymbol{\chi }_{t+1}$ at time $t+1$. We determine the size of $\varDelta \boldsymbol{\chi }_{t+1}$ by setting the value of $\delta$. Then, we obtain $\varDelta \boldsymbol{\chi }_t$ at consecutive times by following the same process.
\subsection{Determine Protection Location Set}
After obtaining $\varDelta \boldsymbol{\chi }_t$ for each time on the trajectory, we consider the protection of possible locations within $\varDelta \boldsymbol{\chi }_t$ at any given time.

In order to improve the user's QoS, the smaller the diameter $D\left( \boldsymbol{\Phi }_t \right)$ of the circular area, the better. Since $D\left( \boldsymbol{\Phi }_t \right)$ is the diameter of the $\boldsymbol{\Phi }_t$, the distance between any two locations is less than or equal to $D\left( \boldsymbol{\Phi }_t \right)$. For $\forall \boldsymbol{x}_t,\boldsymbol{\hat{x}}_t$ in  $\boldsymbol{\Phi }_t$, we have $D\left( \boldsymbol{\Phi }_t \right) \ge d\left( \boldsymbol{x}_t,\boldsymbol{\hat{x}}_t \right)$. By (\ref{Constraint condition}), we have
\begin{small}
\begin{equation}\label{24}
e^{\epsilon}E_m\le E\left( \boldsymbol{\Phi }_t \right) \le \underset{\boldsymbol{\hat{x}}_t\in \boldsymbol{\Phi }_t}{\min}\sum_{\boldsymbol{x}_t\in \boldsymbol{\Phi }_t}{\frac{\text{Pr}\left( \boldsymbol{x}_t \right)}{\sum_{\boldsymbol{y}_t\in \boldsymbol{\Phi }_t}{\text{Pr}\left( \boldsymbol{y}_{\boldsymbol{t}} \right)}}}D\left( \boldsymbol{\Phi }_t \right) =D\left( \boldsymbol{\Phi }_t \right).
\end{equation}
\end{small}
To effectively find the PLS with the smallest diameter at time $t$, the search method based on the Hilbert curve in \cite{yu2017dynamic} is adopted. For each possible location $\boldsymbol{x}_t$ in $\varDelta \boldsymbol{\chi }_t$ on the trajectory, we search the neighborhood of $\boldsymbol{x}_t$ according to the search direction of the Hilbert curve. We identify the PLS for $\boldsymbol{x}_t$  that satisfies (\ref{Constraint condition}) and select the one with the smallest diameter as the PLS $\boldsymbol{\Phi }_t$.

On this basis, to prevent the single-direction search of the Hilbert curve might lead to an unreasonable protection area with a large diameter, we perform spatial rotation of the Hilbert curve to improve the opportunity of finding a PLS for each location $\boldsymbol{x}_t$ with a smaller diameter.
More specifically, similar to \cite{yu2017dynamic} we rotate 90, 180, and 270 degrees clockwise around the center point to generate three more Hilbert curves. After rotation, search for PLS where the user's location is under different Hilbert curves. Then, the group with the smallest diameter is selected from the four results as the PLS.
\subsection{Differentially Private Mechanism in Protection Location Set}
We put forth a new perturbation mechanism, Permute-and-Flip, to release the perturbed location with a smaller perturbation distance, which can better balance location privacy
and QoS. The Permute-and-Flip mechanism was initially developed to protect privacy in the data publishing process \cite{mckenna2020permute}. We apply this mechanism for the first time to protect the location in the PLS $\boldsymbol{\Phi }_t$ through the mapping relationship between the utility function and the Euclidean distance. The Permute-and-Flip mechanism always selects the query option with the highest score when processing query options. Therefore, we take the difference between its distance and the maximum distance as a query function and define the sensitivity of the utility function as
\begin{equation}\label{25}
\varDelta u=\underset{\boldsymbol{x}_{t}^{'}\in \mathcal{A},\boldsymbol{x}_t,\boldsymbol{y}_t\in \boldsymbol{\Phi }_t}{\max\max} \left|d\left( \boldsymbol{x}_t,\boldsymbol{x}_{t}^{'} \right) -d\left( \boldsymbol{y}_t,\boldsymbol{x}_{t}^{'} \right)\right|,
\end{equation}
according to the triangle inequality, we have $\left|d\left( \boldsymbol{x}_t,\boldsymbol{x}_{t}^{'} \right) -d\left( \boldsymbol{y}_t,\boldsymbol{x}_{t}^{'} \right) \right|\le d\left( \boldsymbol{x}_t,\boldsymbol{y}_t \right) \le D\left( \boldsymbol{\Phi }_t \right)$.

After obtaining  $\varDelta \boldsymbol{\chi }_t$ for each location on the trajectory, we can find the corresponding $\boldsymbol{\Phi }_t$ for each possible location in $\varDelta \boldsymbol{\chi }_t$ using (\ref{Constraint condition}). Given the current location $\boldsymbol{x}_t$ and the PLS $\boldsymbol{\Phi }_t$, the probability of the output perturbed location $\boldsymbol{x}_{t}^{'}$ is proportional to $\exp\left( \frac{-\epsilon \left( u\left( D,r \right) -\max\left( u\left( D,r \right) \right) \right)}{2\varDelta u} \right)$ according to the Permute-and-Flip mechanism. We have the perturbed locations' probability distribution
\begin{footnotesize}
\begin{equation}\label{25}
f\left( \boldsymbol{x}_{t}^{'}|\boldsymbol{x}_t \right) =\omega _x\exp\left( \frac{-\epsilon \left( d\left( \boldsymbol{x}_t,\boldsymbol{x}_{t}^{'} \right) -\max d\left( \boldsymbol{x}_t,\boldsymbol{x}_{t}^{'} \right) \right)}{2D\left( \boldsymbol{\Phi }_t \right)} \right),
\end{equation}
\end{footnotesize}where $\omega _x$ is the probability distribution normalization factor, i.e.,
\begin{scriptsize}
\begin{equation}\label{26}
\omega _x=\left( \sum_{\boldsymbol{x}_{t}^{'}\in \mathcal{A}}{\exp \left( \frac{-\epsilon \left( d\left( \boldsymbol{x}_t,\boldsymbol{x}_{t}^{'} \right) -\max d\left( \boldsymbol{x}_t,\boldsymbol{x}_{t}^{'} \right) \right)}{2D\left( \Phi _t \right)} \right)} \right) ^{-1}.
\end{equation}
\end{scriptsize}
\begin{figure}[!t]

\begin{center}

\includegraphics[width=0.4\textwidth]{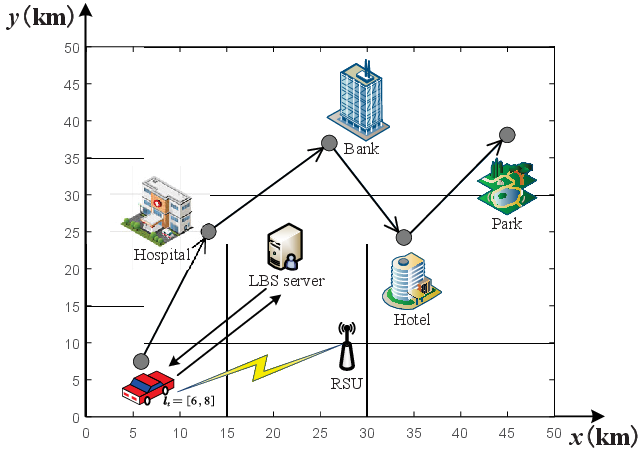}

\end{center}

\caption{Simulation setting of the trajectory of a user.}

\label{fig4}

\end{figure}
\section{Simulation Results}\label{Simulation Results}
In this section, we evaluate the effectiveness of our proposed PTPPM. We compare the trajectory privacy performance of PTPPM with that of PIVE \cite{yu2017dynamic} under the same QoS loss. To facilitate evaluation, we divide the $50\ \text{km}\times 50\ \text{km}$ two-dimensional space evenly into 100 units, and each unit has the same area.
These units serve as areas that VANET users may access, also known as the attacker's prior distribution. Each unit represents the location status of the user and has corresponding two-dimensional coordinates. We select 5 of them as the real locations of 5 consecutive moments on the user's trajectory, as depicted in Fig. \ref{fig4}.

The location privacy $p$ and QoS loss $q$ are evaluated by the similar metrics in our previous work \cite{min20213d} which are given by
\begin{equation}\label{Privacy}
p=\sum_{\boldsymbol{x}_t,\boldsymbol{x}_{t}^{'},\boldsymbol{\hat{x}}_t\in \mathcal{A}}{\text{Pr}\left( \boldsymbol{x}_t \right)}f\left( \boldsymbol{x}_{t}^{'}|\boldsymbol{x}_t \right) h\left( \boldsymbol{\hat{x}}_t|\boldsymbol{x}_{t}^{'} \right) d\left( \boldsymbol{x}_t,\boldsymbol{\hat{x}}_t \right),
\end{equation}
\begin{equation}\label{QoS loss}
q=\sum_{\boldsymbol{x}_t,\boldsymbol{x}_{t}^{'}\in \mathcal{A}}{\text{Pr}\left( \boldsymbol{x}_t \right)}f\left( \boldsymbol{x}_{t}^{'}|\boldsymbol{x}_t \right) d\left( \boldsymbol{x}_t,\boldsymbol{x}_{t}^{'} \right).
\end{equation}

\begin{figure}[!t]
\centering
\subfigure[Trajectory privacy v.s. $\epsilon$]{
\includegraphics[height=3cm,width=3.9cm]{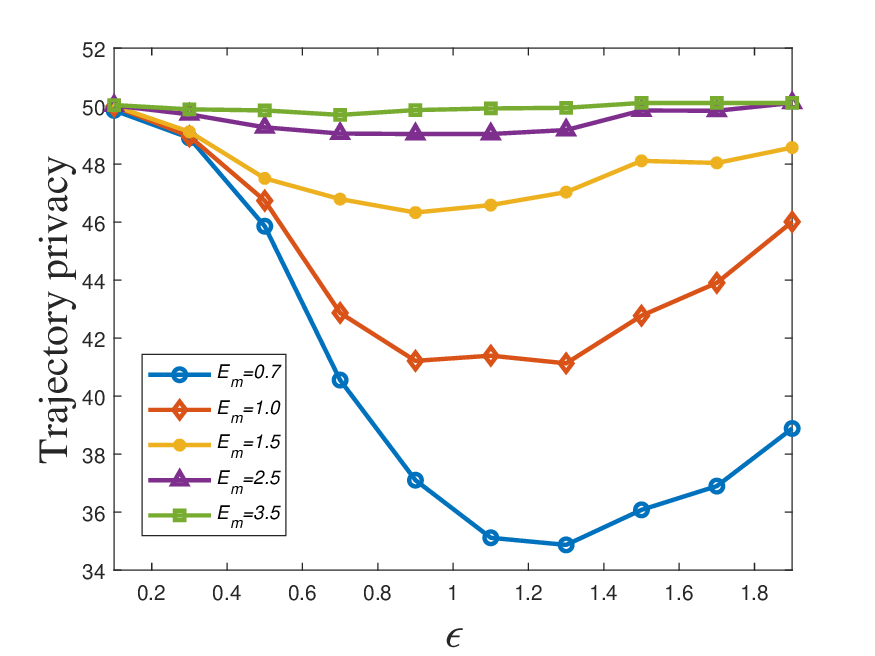} \label{fig5a}
}
\quad
\subfigure[Trajectory privacy v.s. $E_m$]{
\includegraphics[height=3cm,width=3.9cm]{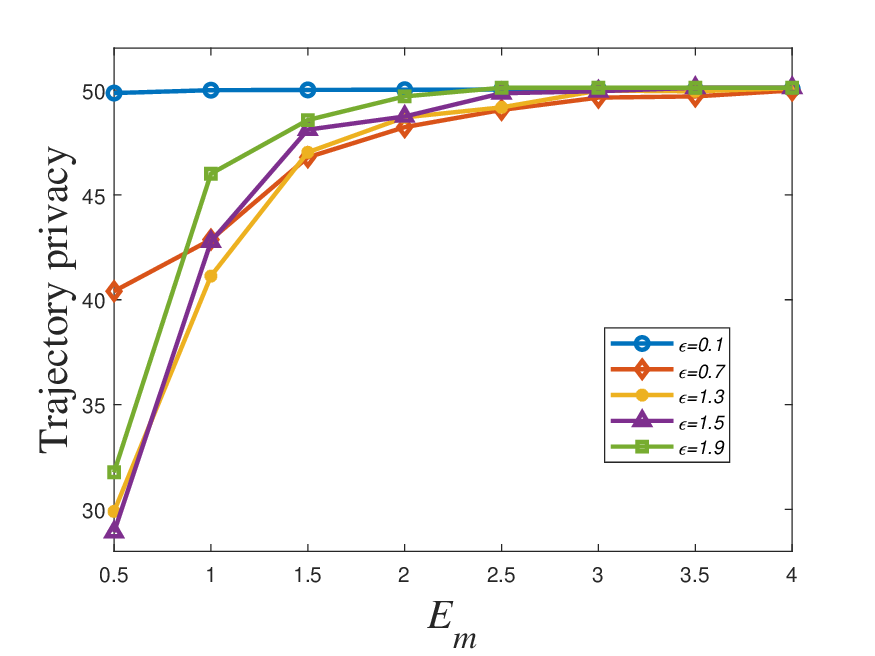} \label{fig5b}
}
\quad
\subfigure[QoS loss v.s. $\epsilon$]{
\includegraphics[height=3cm,width=3.9cm]{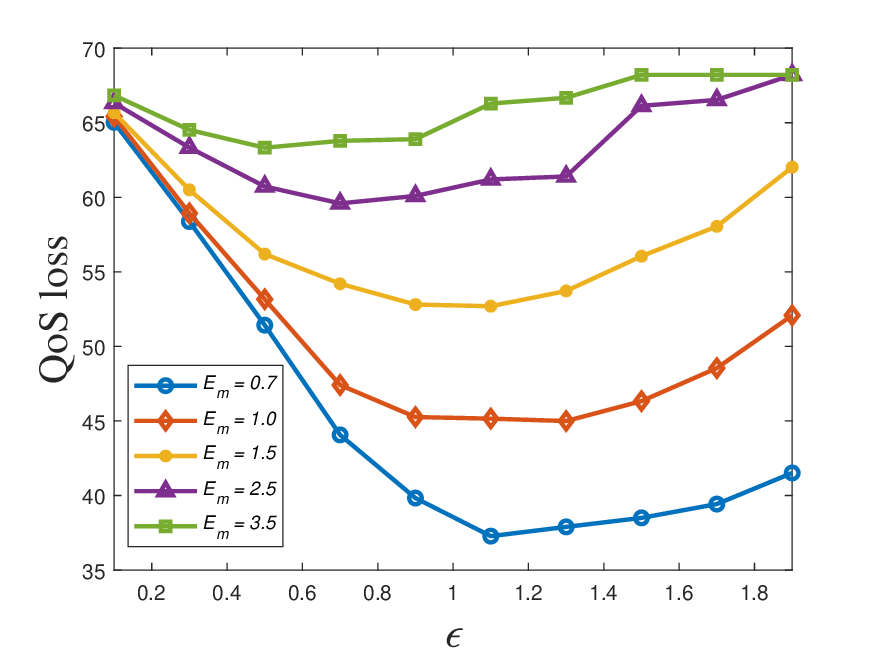}\label{fig5c}
}
\quad
\subfigure[QoS loss v.s. $E_m$]{
\includegraphics[height=3cm,width=3.9cm]{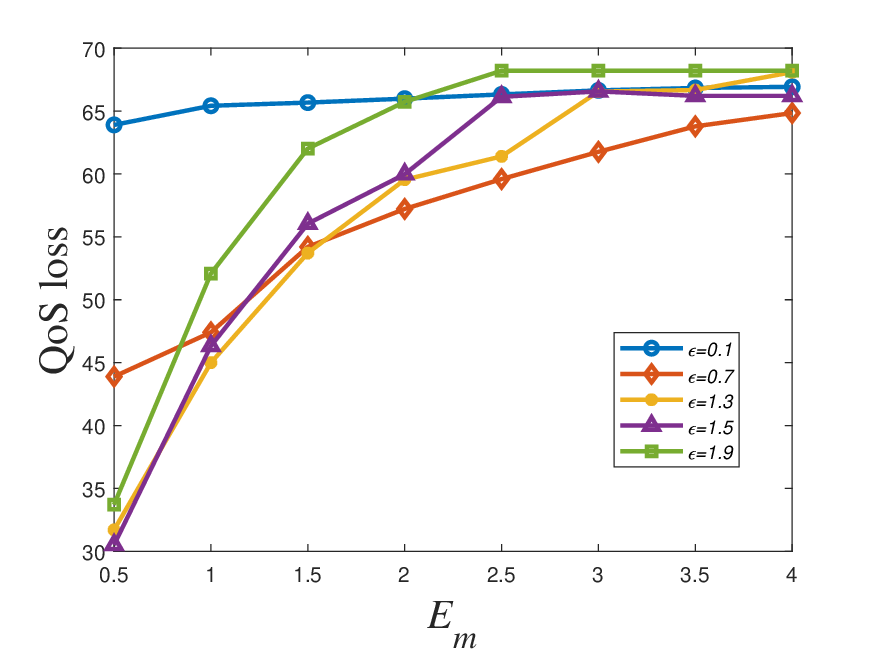}\label{fig5d}
}
\caption{Impact of $\epsilon$ and $E_m$ on personalized trajectory privacy protection.}
\end{figure}

First, we set different privacy budgets $\epsilon$ and inference error threshold $E_m$ to evaluate their impact on users' personalized trajectory privacy protection performance in Figs. 5. We can see that two privacy parameters ($\epsilon$ and $E_m$) have a significant impact on trajectory privacy and QoS loss. More specifically, as shown in Figs. \ref{fig5a} and \ref{QoS_epsilon}, when $\epsilon$ is small, the trajectory privacy and QoS loss decrease with increased $\epsilon$ under different $E_m$. Besides, when $\epsilon$ is larger than a specific value, the trajectory privacy and QoS loss start to increase. That is because, according to (\ref{Constraint condition}), the increase of $\epsilon$ will cause $D\left( \boldsymbol{\Phi }_t \right)$ to sharply increase. The turning points under different $E_m$ settings are different. Moreover, because $D\left( \boldsymbol{\Phi }_t \right)$ cannot be increased indefinitely in practical scenarios, the location privacy and QoS loss finally reach the upper limit value.
Figs. \ref{fig5b} and \ref{fig5d} show that the trajectory privacy, QoS loss, and trajectory error increase with the increase of $E_m$ under different $\epsilon$. Given a $\epsilon$, when $E_m$ increases, the $D\left( \boldsymbol{\Phi }_t \right)$ of the protected area increases, thus increasing the trajectory privacy and QoS loss. Moreover, the effect of $\epsilon$ on $D\left( \boldsymbol{\Phi }_t \right)$ is exponential, much higher than that of $E_m$. Therefore, when $\epsilon$ is set to 1.5, with the increase of $E_m$, $D\left( \boldsymbol{\Phi }_t \right)$ changes significantly, so the trajectory privacy and QoS loss increase sharply, resulting in a steep curve. However, there is a limitation on the $D\left( \boldsymbol{\Phi }_t \right)$ of the protected area, so the trajectory privacy and QoS loss will converge to a finite value. We can see that the user's trajectory privacy and QoS loss reach the upper limit when $\epsilon$ is 0.1, regardless of the $E_m$ setting. By adjusting different privacy settings, personalized trajectory privacy protection is realized.

\begin{figure}[!t]

\begin{center}

\includegraphics[width=0.37\textwidth]{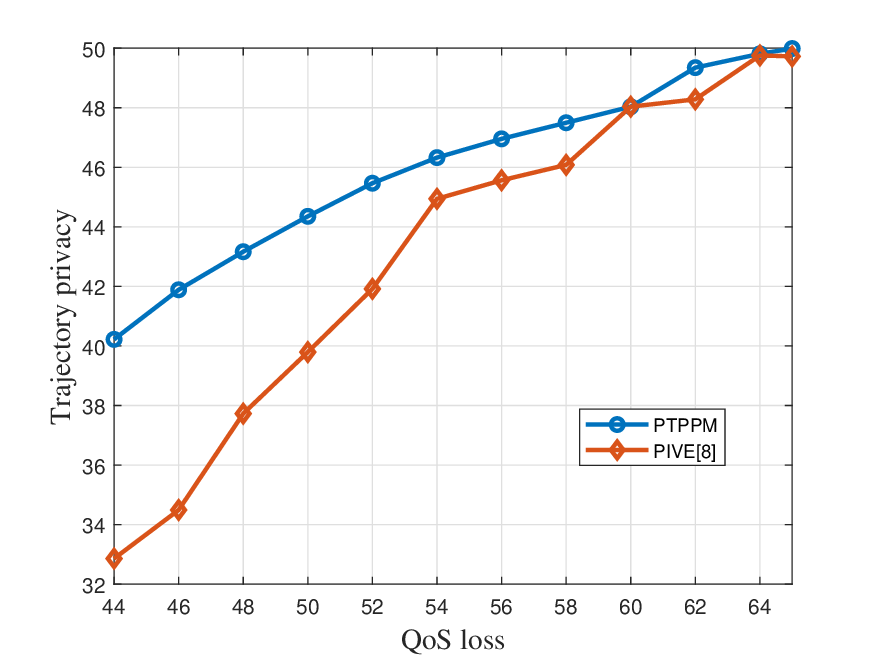}

\end{center}

\caption{Performance of different LPPMs under different QoS loss.}

\label{fig6}

\end{figure}

Next, we quantitatively compare PF with PIVE in terms of trajectory privacy and QoS loss to verify its advantages. We set (\ref{QoS loss}) equal to the set QoS loss value, and the only variable in this equation is the privacy budget $\epsilon$. By solving this equation, $\epsilon$ corresponding to PF and PIVE can be obtained under the same QoS loss. By substituting (\ref{Privacy}), the corresponding privacy of PF and PIVE under the same QoS loss can be calculated. As shown in Fig. \ref{fig6}, we can see that PF can better protect privacy under the same QoS loss. For example, when QoS loss = 44, the privacy value of PTPPM is 22.4\% which is higher than that of PIVE. That is because the proposed Permute-and-Flip mechanism provides a smaller perturbation distance while guaranteeing privacy demands in PLS. In addition, since $D\left( \boldsymbol{\Phi }_t \right)$ cannot be infinitely enlarged in the actual scenario, privacy eventually reaches the upper limit. We can see that the proposed mechanism can better protect user privacy while meeting users' QoS requirements.
\section{Conclusion}\label{Conclusion}
In this paper, we have proposed a personalized trajectory privacy protection mechanism PTPPM. This paper has three novel contributions: First, we address the issue of attackers exploiting the temporal correlation between different locations to compromise user privacy. To mitigate this threat, we design a robust trajectory privacy protection mechanism. Second, we combined the privacy notions of geo-indistinguishability and distortion privacy, enabling personalized privacy protection by adjusting the privacy budget and the expected inference error threshold to meet individual user needs. Third, we proposed a novel perturbation mechanism, Permute-and-Flip, which releases perturbed locations with smaller perturbation distances to better balance the trajectory privacy and QoS. Simulation results show that PTPPM offers improved privacy protection under the same QoS loss compared to PIVE. For instance, when QoS loss = 44, the privacy of PTPPM is 22.4\% higher than that of PIVE.

\bibliography{ref}
\bibliographystyle{ieeetr}

\end{document}